# FPGA Implementation of Pipeline Digit-Slicing Multiplier-Less Radix $2^2$ DIF SDF Butterfly for Fast Fourier Transform Structure

[1]Yazan Samir Algnabi, [1,2]Rozita Teymourzadeh, [1]Masuri Othman, [1]Md Shabiul Islam

[1]Institute of MicroEngineering and Nanoelectronics IMEN, VLSI Design Department,
Universiti Kebangsaan Malaysia, 43600 Bangi, Selangor, Malaysia

[2]Faculty of Engineering, Architecture and Built Environment,
Electrical & Electronic Engineering department, UCSI University, Kuala Lumpur, Malaysia

yazansamir@yahoo.com, rozita@ucsi.edu.my, masuri.othman@mimos.my, shabiul@ukm.my

*Abstract*—The need for wireless communication has driven the communication systems to high performance. However, the main bottleneck that affects the communication capability is the Fast Fourier Transform (FFT), which is the core of most modulators. This paper presents FPGA implementation of pipeline digit-slicing multiplier-less radix $2^2$ DIF (Decimation In Frequency) SDF (single path delay feedback) butterfly for FFT structure. The approach taken; in order to reduce computation complexity in butterfly multiplier, digit-slicing multiplier-less technique was utilized in the critical path of pipeline Radix-$2^2$ DIF SDF FFT structure. The proposed design focused on the trade-off between the speed and active silicon area for the chip implementation. The multiplier input data was sliced into four blocks each one with four bits to process at the same time in parallel. The new architecture was investigated and simulated with MATLAB software. The Verilog HDL code in Xilinx ISE environment was derived to describe the FFT Butterfly functionality and was downloaded to Virtex II FPGA board. Consequently, the Virtex-II FG456 Proto board was used to implement and test the design on the real hardware. As a result, from the findings, the synthesis report indicates the maximum clock frequency of 555.75 MHz with the total equivalent gate count of 32,146 is a marked and significant improvement over Radix $2^2$ DIF SDF FFT butterfly. In comparison with the conventional butterfly architecture design which can only run at a maximum clock frequency of 200.102 MHz and the conventional multiplier can only run at a maximum clock frequency of 221.140 MHz, the proposed system exhibits better results. It can be concluded that on-chip implementation of pipeline digit-slicing multiplier-less butterfly for FFT structure is an enabler in solving problems that affect communications capability in FFT and possesses huge potentials for future related works and research areas.

*Key words*— Pipelined digit-slicing multiplier-less; Fast Fourier Transform (FFT); Verilog HDL; Xilinx; Radix 22 DIF SDF FFT.

## I. INTRODUCTION

FFT is significant block in several digital signal processing (DSP) applications such as biomedical, sonar, communication systems, radar, and image processing. It is a successful algorithm to compute discrete Fourier transform (DFT). DFT is the main and important procedure in data analysis, system design, and implementation [1]. Many modules have been designed and implemented in different platforms in order to reduce the complexity computation of the FFT algorithm. These modules focus on the radix order or twiddle factors to perform a simple and efficient algorithm which includes the higher radix FFT [2], the mixed-radix FFT [3], the prime-factor FFT [4], the recursive FFT [5], low-memory reference FFT [6], Multiplier-less based FFT [7, 8] and Application-Specific Integrated Circuits (ASIC) system [9, 10]. A special class of FFT architecture which can compute the FFT in a sequential manner is the pipeline FFT. Pipelined architectures characterized by real-time, non stopping processing and present smaller latency with low power consumption [11] which makes them suitable for most DSP applications. There are two common types of the pipelined architectures; single path architectures and multi path architectures. Several different architectures have been investigated, such as the Radix 2 Multi-path Delay Commutator (R2 MDC) [12], Radix 2 Single-Path Delay Feedback (R2 SDF) [13], Radix 4 Single-Path Delay Commutator (R4 SDC) [14], and Radix-$2^2$ Single-Path Delay Feedback (R$2^2$ SDF) [15]. The study made on the listed architectures shows that the Delay Feedback architecture is more efficient than the other delay commutator in terms of memory utilization. Radix-$2^2$ has simpler butterfly as Radix 2 and the same multiplicative complexity as Radix 4 algorithm [16, 17]. This makes Radix-$2^2$ single path delay feedback an attractive architecture for DSP implementation. The study of the digit-slicing technique has been dealt by [18-20] for the digital filters. The design and implementation of Digit-slicing FFT has been discussed by [21]. This paper proposed a similar idea with the ones put forth by [21]; but having a difference by the use of a different algorithm, structure and different platform, which helps to improve the performance and achieve higher clock frequency. Recently, Field Programmable Gate Array (FPGA) has become an applicable option to direct hardware solution performance in the real time application. In this paper, digit-slicing architecture is proposed to design the pipeline digit-slicing multiplier-less Radix $2^2$ SDF butterfly. The FFT butterfly multiplication is the most crucial part in causing the delay in the computation of the FFT. In view of the fact, the twiddle factors in the FFT processor were known in advance hence we proposed to use the pipeline digit slicing multiplier-less butterfly to replace the traditional butterfly in FFT.

## II. RADIX $2^2$ SDF FFT ALGORITHM

The more efficient architecture in terms of memory utilization is the delay feedback. radix-4 algorithm based single-path architectures have higher multiplier utilization; however, radix-2 algorithm based architectures have simpler butterflies and control logic. The radix $2^2$ FFT algorithm has the same multiplicative complexity as radix 4 but retains the butterfly structure of radix 2 algorithm [15]. That makes the R2$^2$ SDF FFT algorithm the best choice for the VLSI implementation. In this algorithm, the first two steps of the decomposition of radix 2 DIT FFT are analysed, and common factor algorithm is used to illustrate.

$$X[k] = \sum_{n=0}^{N-1} x[n] W_N^{nk}, k = 0,1,\ldots,N-1 \quad (1)$$

$x[n]$ and $X[k]$ = Complex numbers
$W_N^{kn} = e^{-j2\pi/N}$ = The twiddle factor

In the Equation (1) the index $n$ and $k$ decomposed as:

$$n = <\frac{N}{2} n_1 + \frac{N}{4} n_2 + n_3 >_N \quad (2)$$

$$k = <k_1 + 2k_2 + 4k_3 >_N \quad (3)$$

The total value of n and k is N. when the above substations are applied to Equation (1) the DFT definition can be written as:

$$X[k_1 + 2k_2 + 4k_3] =$$
$$\sum_{n_3=0}^{(N/4)-1} \sum_{n_2=0}^{1} \sum_{n_1=0}^{1} x\left[\frac{N}{2}n_1 + \frac{N}{4}n_2 + n_3\right] W_N^{\left(\frac{N}{2}n_1 + \frac{N}{4}n_2 + n_3\right)(k_1 + 2k_2 + 4k_3)}$$

$$X[k_1 + 2k_2 + 4k_3] =$$
$$\sum_{n_3=0}^{(N/4)-1} \sum_{n_2=0}^{1} \left[ B_{N/2}^{k_1}\left[\frac{N}{4}n_2 + n_3\right] W_N^{\left(\frac{N}{4}n_2 + n_3\right)} \right] W_N^{\left(\frac{N}{4}n_2 + n_3\right)(2k_2 + 4k_3)}$$

Where:

$$B_{N/2}^{k_1} = x\left[\frac{N}{4}n_2 + n_3\right] + (-1)^{k_1} x\left[\frac{N}{4}n_2 + n_3 + \frac{N}{2}\right] \quad (6)$$

For normal radix 2 DIF FFT algorithms, the expression in the braces is computed first as a first stag in Equation (5). However, in radix $2^2$ FFT algorithm, the main idea is to reconstruct the first stage and the second stage twiddle factors, as shown in Equation (8) as mentioned in [15].

$$W_N^{\left(\frac{N}{4}n_2+n_3\right)k_1} \quad W_N^{\left(\frac{N}{4}n_2+n_3\right)(2k_2+4k_3)} = \quad (7)$$
$$W_N^{Nn_2k_3} W_N^{Nn_2(k_1+2k_2)} W_N^{n_3(k_1+2k_2)} W_N^{4n_3k_3}$$

$$W_N^{\left(\frac{N}{4}n_2+n_3\right)k_1} \quad W_N^{\left(\frac{N}{4}n_2+n_3\right)(2k_2+4k_3)} = \quad (8)$$
$$(-j)^{n_2(k_1+2k_2)} W_N^{n_3(k_1+2k_2)} W_N^{4n_3k_3}$$

Observe that the last twiddle factor in Equation (8) can be rewritten as:

$$W_N^{4n_3k_3} = e^{\frac{-j2\pi}{N}(4n_3k_3)} = e^{\frac{-j2\pi}{4N}(n_3k_3)} = W_{N/4}^{n_3k_3} \quad (9)$$

By applying Equation (8) and (9) in Equation (5) and expand the summation over $n_2$, the result is a DFT definition with four times shorter FFT length.

$$X[k_1 + 2k_2 + 4k_3] = \\ \sum_{n_3=0}^{(N/4)-1} \left[ H[k_1 + 2k_2 + 4k_3] W_N^{n_3(k_1+2k_2)} \right] W_N^{n_3k_3} \quad (10)$$

Where,

$$H[k_1 + 2k_2 + 4k_3] = \left[ x(n_3) + (-1)^{k_1} x\left(n_3 + \frac{N}{2}\right) \right] \\ + (-j)^{(k_1+2k_2)} \left[ x\left(n_3 + \frac{N}{4}\right) + (-1)^{k_1} x\left(n_3 + \frac{3N}{4}\right) \right] \quad (11)$$

Each term in equation (10) represents a Radix-2 butterfly (Butterfly I), while the whole equation represents Radix-2 butterfly, (Butterfly II) with trivial multiplication by (-j). Equation (10) known as radix $2^2$ SDF FFT algorithm. Fig. 1 shows the butterfly signal flow graph for radix $2^2$ FFT algorithm. Fig. 2 shows the 16 point R2$^2$ SDF FFT signal flow graph.

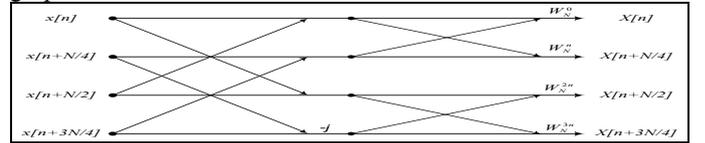

Fig. 1 The butterfly structure for the radix $2^2$ DIF FFT

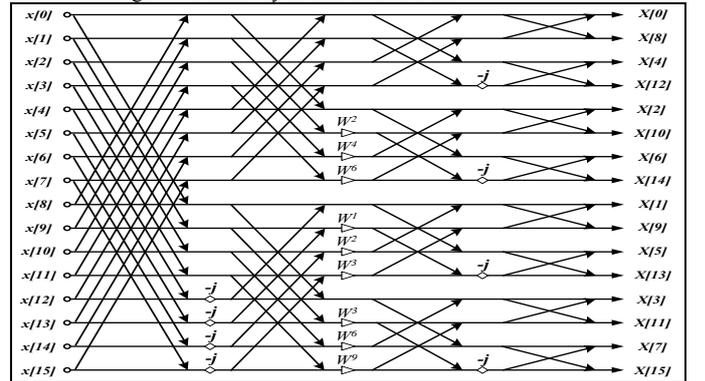

Fig. 2 16-Point R2$^2$ SDF DIF FFT signal flow graph.

## III. RADIX $2^2$ SDF BUTTERFLY STRUCTURE

From equation (10), each stage in R2$^2$ SDF FFT consists of Butterfly I, Butterfly II, Complex multipliers with twiddle factors. Butterfly I calculate the input data flow, butterfly II calculate the output data flow from Butterfly I, than multiply the twiddle factors with the output data from Butterfly II, to get the result of the current stage. Fig. 3 shows the structure of 16 point R2$^2$ SDF FFT.

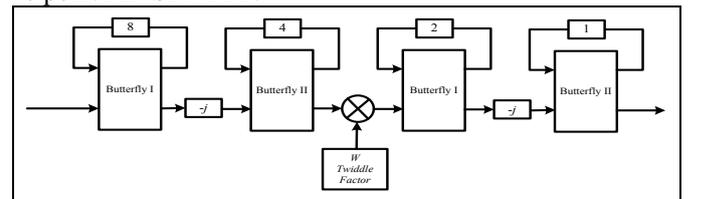

Fig. 3 16-Point R2$^2$ SDF DIF FFT Structure.

## A. Butterfly I Structure

Fig. 4 shows the Butterfly I structure, the input $A_r$, $A_i$ for this butterfly comes from the previous component which is the twiddle factor multiplier except the first stage it comes form the FFT input data. The output data $B_r$, $B_i$ goes to the next stage which is normally the Butterfly II. The control signal C1 has two options C1=0 to multiplexers direct the input data to the feedback registers until they filled. The other option is C1=1 the multiplexers select the output of the adders and subtracters.

The process of the Butterfly I is to store the anterior half of the N point input series in feedback registers, than butterfly calculation when the posterior half data is coming, the result of the butterfly is $B_r$, $B_i$, $D_r$, $D_i$. $B_r$, $B_i$ fed to the output result of the Butterfly I the other result $D_r$, $D_i$ goes to the feedback registers.

## B. Butterfly II Structure

Fig. 5 shows the Butterfly II structure b. The input data $B_r$, $B_i$ comes from the previous component, Butterfly I. The output data from the Butterfly II are $E_r$, $E_i$, $F_r$ and $F_i$. $E_r$, $E_i$ fed to the next component, normally twiddle factor multiplier. The $F_r$ and $F_i$ go to the feedback registers.

The multiplication by $-j$ involves swapping between real part and imaginary part and sign inversion. The swapping is handled by the multiplexers Swap-MUX efficiently and the sign inversion is handled by switching between the adding and the subtracting operations by mean of Swap-MUX. The control signals C1 and C2 will be one when there is a need for multiplication by $-j$, therefore the real and imaginary data will swap and the adding and subtracting operations will switched.

In order to not lose any precision the divide by 2 is used where the word lengths imply successive growth as the data goes through adder, subtracter and multiplier operations. Rounding off has been also applied to reduce the scaling errors.

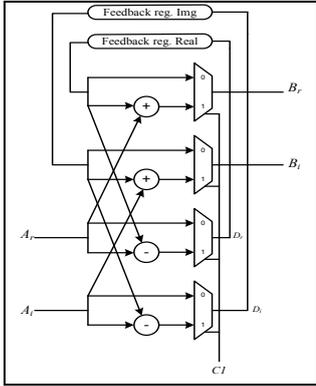
Fig. 4 The Butterfly I Structure.

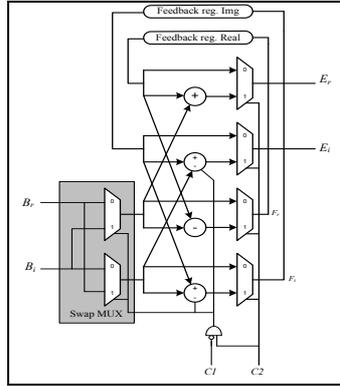
Fig. 5 The Butterfly II Structure.

## C. Complex Multiplier

Normally the complex multiplier can be realized by four real multipliers, one adder and one subtractor as shown in Fig. 6. This complex multiplier structure occupies large chip area in VLSI implementation.

$$(a_r+ja_i)(b_r+jb_i) = (a_rb_r-a_ib_i) + j(a_ib_r+a_rb_i) \quad (12)$$

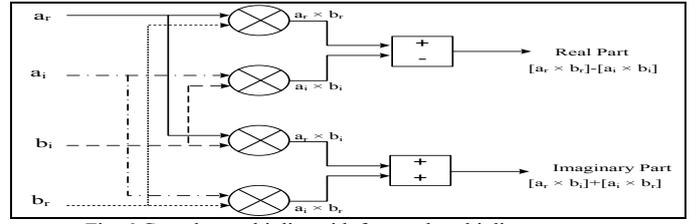
Fig. 6 Complex multiplier with four real multiplier structure.

This complex multiplier can be realized by only three real multipliers and five real adder/ subtractor based on equation (13); this will save a lot of area in hardware implementation as shown in Fig. 7.

$$(a_r+ja_i)(b_r+jb_i)=\{b_r(a_r-a_i)+a_i(b_r-b_i)\}+j\{b_i(a_r+a_i)+a_i(b_r-b_i)\} \quad (13)$$

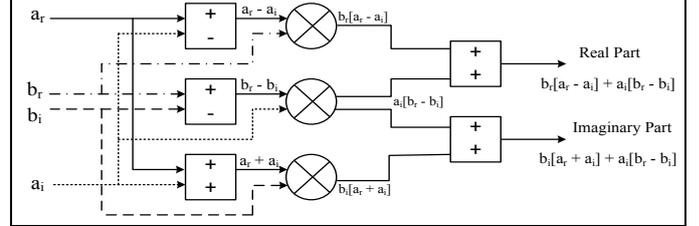
Fig. 7 Complex multiplier with three real multiplier structure.

## IV. FPGA IMPLEMENTATION OF PIPELINE DIGIT-SLICING MULTIPLIER-LESS RADIX $2^2$ DIF SDF BUTTERFLY

Previous section explain in details the conventional structure of the R$2^2$ SDF butterfly, this section discuss how to apply the digit slicing technique for the R$2^2$ SDF butterfly component in order to reduce the complexity computation and enhanced the throughput.

The digit slicing multiplier less R$2^2$ SDF butterfly has been used the same component of the conventional structure except the complex multiplier which has been replaced with the digit slicing multiplier less.

The multiplication functionality is regarded as the most important operation for most signal processing systems, but it is a complex and expensive operation. Many techniques have been introduced for reducing the size and improving the speed of multipliers. In this paper we proposed digit slicing multiplier less to improve the speed of the multiplication. The design of the digit slicing complex multiplier has been made by Matlab to prove the working of the algorithm than we improved the design to be the digit slicing multiplier less.

The concept behind the digit slicing architecture is any binary number can be sliced into a few blocks of shorter binary numbers, with each block carrying a different weight [22]. In this paper, the 16 bits fixed-point 2's complements arithmetic has been chosen to represent the input data and the twiddle factor, which are singed numbers with absolute value less than one. Let us conceder the absolute value of the complex multiplier input data (the output of Butterfly II) is $x$ with length of 16 bits has been represented in 2's complement as:

$$x = \sum_{k=0}^{B-1} 2^{-j} x^j \quad (13)$$

To represent the sliced data, the fundamental sliced algorithm will be presented as following:

$$x = \left[\sum_{k=0}^{b-1} 2^{pk} X_k\right] 2^{-(pb-1)} \qquad X_k = \sum_{j=0}^{p-1} 2^j X_{k,j} \qquad (14)$$

Where $x$ is sliced into b blocks and p is bit widths per block and $X_{k,j}$ are all either ones or zeros except for $X_k$=b-1, j=p-1 which is zero or minus one. The digit slicing architecture has been applied for the complex multiplier input data (the output of Butterfly II) to slice the data to four groups each carrying four bits as shown in Fig. 8 and Fig. 9.

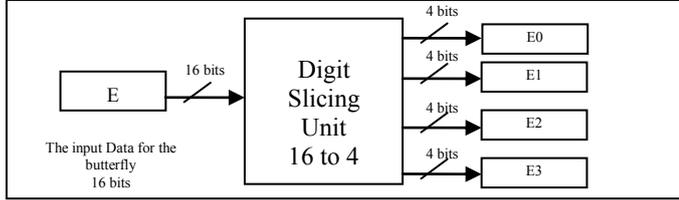
Figure 8. Digit Slicing Structure

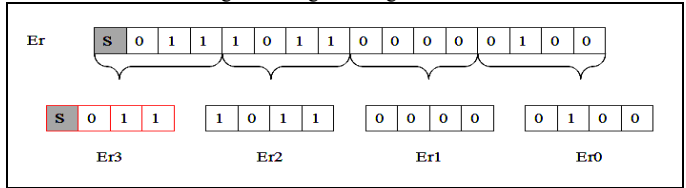
Figure 9. Digit Slicing for the input $E_r$

The complex multiplier realized by three real multipliers, as mention in previous section the digit slicing has been applied for the real multiplier input data to make the multiplication process parallel with the 16 bits twiddle factor as shown in Fig. 10. Therefore the processing time will be reduced. To understand and prove the digit slicing algorithm the MATLAB design for the complex multiplier and the digit slicing multiplier has been made and the result has been compared as shown in Fig. 11 and Fig. 12.

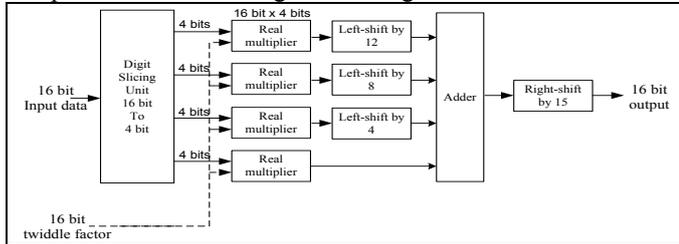
Fig. 10 digit slicing multiplier structure.

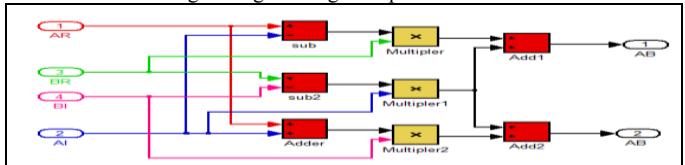
Fig. 11 Design of Complex multiplier in MATLAB.

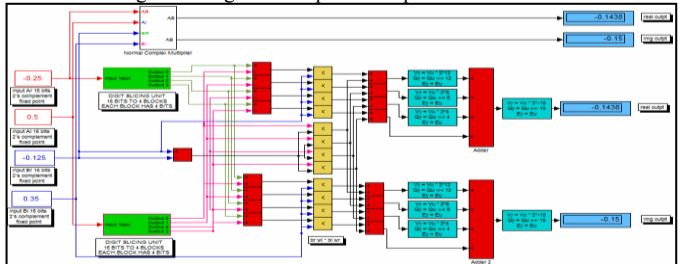
Fig. 12 Design of digit slicing Complex multiplier in MATLAB.

Since the twiddle factors in FFT are known in advanced therefore the multiplication possibility for the 16 bits twiddle factor and multiply by 4 bits input data will be 16 possibilities can be stored in one RAM for each twiddle factor. This design will improve the digit slicing multiplier to be digit slicing multiplier less which has been replaced with the conventional multiplier as shown in Fig. 13.

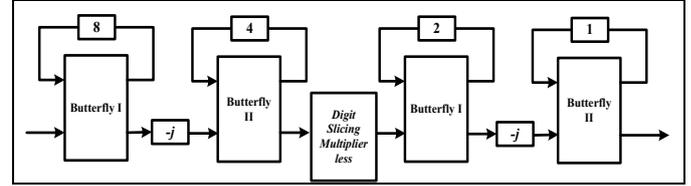
Fig. 13 16-point R22 SDF FFT structure with digit slicing multiplier

The design of the digit slicing multiplier less consists of one lookup table (ROM) shift and adder to perform the output as shown in Fig 14. and Fig. 15. To generate the lookup table data (the multiplication result possibilities), which are 16 different results, a special MATLAB program has been written by applying the digit-slicing algorithm for all the possible numbers for the input data (4 bits) from "0000" to "1111" to perform all the possibilities for the multiplication result. The storage of all these possibilities in one ROM allows the design to perform the multiplication process without any real multiplier.

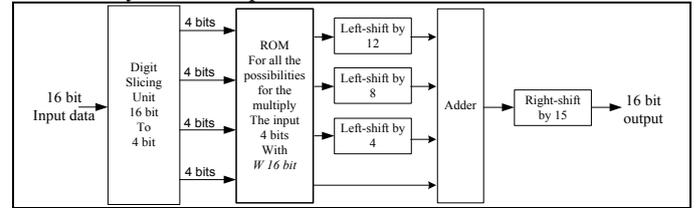
Fig. 14 digit slicing multiplier less structure

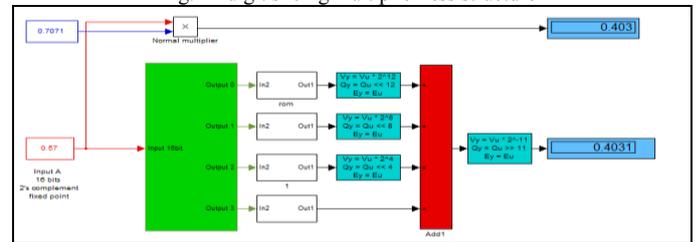
Fig. 15 design of digit slicing multiplier less in MATLAB

The Verilog HDL code in Xilinx ISE environment was derived to describe the Pipeline Digit-Slicing Multiplier-Less Radix $2^2$ DIF SDF Butterfly functionality and was downloaded to Virtex II FPGA board. Consequently, the Virtex-II FG456 Proto board was used to implement and test the design on the real hardware.

## V. RESULT

Two different modules were implemented for $R2^2$ SDF DIF FFT butterfly. The first module uses the conventional architecture for the butterfly where the twiddle factors are stored in ROM and called by the butterfly to be multiplied with the inputs by utilising the dedicated high speed multiplier equipped with the Virtex-II FPGA.

The other module uses the pipelined digit-slicing multiplier-less architecture to perform the multiplication with the twiddle factor. Both modules were built and tested in MATLAB as indicated in previous section, then coded in Verilog and synthesized by using the XST-Xilinx Synthesis Technology tool. The target FPGA was Xilinx Virtex-II XC2V500-6-FG456 FPGA. The ModelSim simulation result of Pipeline Digit-Slicing Multiplier-Less Radix $2^2$ DIF SDF Butterfly is shown in Fig. 16, while the synthesis results for the two models are presented in Table 1, which demonstrates the hardware specifications for the design. It indicates the maximum clock frequency of 555.75 MHz for Pipeline Digit-Slicing Multiplier-Less Radix $2^2$ DIF SDF Butterfly as well as the Pipelined Digit-slicing Single Multiplier-less for the butterfly with a performance of the maximum clock frequency of 609.980 MHz. Meanwhile, Fig. 17 shows the RTL schematic for the Pipeline Digit-Slicing Multiplier-Less Radix $2^2$ DIF SDF Butterfly.

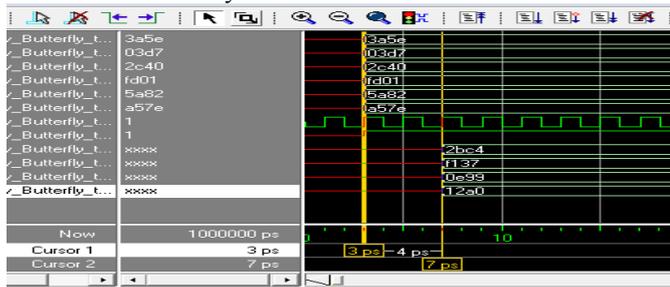

Fig. 16 ModelSim simulation result of Pipeline Digit-Slicing Multiplier-Less Radix $2^2$ DIF SDF Butterfly

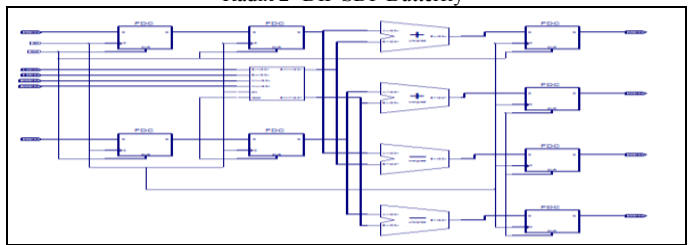

Fig. 17 RTL schematic for the Pipeline Digit-Slicing Multiplier-Less Radix $2^2$ DIF SDF Butterfly

Table 1: Hardware specifications of the digit-slicing butterfly

| Xilinx Virtax-II FPGA XC2v250-6FG456 | Total equivalent gate count for design | Maximum Frq. MHz |
|---|---|---|
| Conventional butterfly | 18.408 | 200.102 |
| Pipeline Digit-Slicing Multiplier-Less Radix $2^2$ DIF SDF Butterfly | 32,146 | 555.75 |
| Conventional 16 bits Multiplier | 4.131 | 220.160 |
| Pipeline Digit-Slicing Multiplier Less 16 bits | 6.483 | 609.980 |

## VI. CONCLUSION

This study presented of FPGA Implementation of Pipeline Digit-Slicing Multiplier-Less Radix $2^2$ DIF SDF Butterfly for FFT Structure. The implementation has been coded in Verilog hardware descriptive language and was tested on Xilinx Virtex-I1 XC2V500-6- FG456 prototyping FPGA board. A maximum clock frequency of 555.75MHz has been obtained from the synthesis report for the Pipeline Digit-Slicing Multiplier-Less Radix $2^2$ DIF SDF Butterfly which is 2.8 time faster than the conventional butterfly. It can be concluded that FPGA Implementation of Pipeline Digit-Slicing Multiplier-Less Radix $2^2$ DIF SDF Butterfly for FFT Structure is an enabler in solving problems that affect communications capability in FFT and possesses huge potentials for future related works and research areas.

REFERENCES

[1] A. V. Oppenheim, R. W. Schafer, and J. R. Buck, Discrete-time signal processing, 2 ed. Upper Saddle River, N.J.: Prentice Hall, 1999.
[2] G. D. Bergland, "A radix-eight fast-Fourier transform subroutine for real-valued series.," IEEE Trans. Audio Electroacoust, vol. 17, pp. 138-144, 1969.
[3] R. C. Singleton, "An algorithm for computing the mixed radix fast Fourier transform," Audio and Electroacoustics, IEEE Transactions on vol. 17, pp. 93-103, 1969.
[4] D. P. Kolba and T. W. Parks, "A prime factor FFT algorithm using high-speed convolution," IEEE Trans Acoust. Speech, Signal Process, vol. 25, pp. 281-294, 1977.
[5] A. R. Varkonyi-Koczy, "A recursive Fast Fourier Transform algorithm," IEEE Trans. Circuits System, vol. 42, pp. 614-616, 1995.
[6] Y. Wang, Y. , Y. J. Tang, J. G. Chung, and S. S. Song, "Novel memory reference reduction methods for FFT implementation on DSP processors," IEEE Trans. Signal Process, vol. 55, pp. 2338-2349, 2007.
[7] Y. Zhou, J. M. Noras, and S. J. Shephend, "Novel design of multiplier-less FFT processors," Signal Proc., vol. 87, pp. 1402-1407, 2007.
[8] B. Mahmud and M. Othman, "FPGA implementation of a canonical signed digit multiplier-less based FFT Processor for wireless communication applications," in ICSE2006 Proc Kuala Lumpur, Malaysia, 2006, pp. 641-645.
[9] B. M. Baas, "An approach to low-power, high-performance fast fourier transform processor design," in Electrical Engineering vol. Ph.D: Stanford University 1999, p. 169.
[10] Y. P. Hsu and S. Y. Lin, "Parallel-computing approach for FFT implementation on Digital Signal Processor (DSP)," World Acad. Sci., Eng. Technol., vol. 42, pp. 587-591, 2008.
[11] T. Sansaloni, A. P´erez-Pascual, V. Torres, and J. Valls, "Efficient pipeline FFT processors for WLAN MIMO-OFDM systems," Electronics Letters, vol. 41, pp. 1043–1044, 2005.
[12] L. R. Rabiner and B. Gold, Theory and application of digital signal processing. Englewood Cliffs, N.J.: Prentice-Hall, 1975.
[13] E. H. Wold and A. M. Despain, "Pipeline and parallel-pipeline FFT processors for VLSI implementation " IEEE Transactions on Computers, vol. 33, pp. 414–426, 1984.
[14] G. Bi and E. V. Jones, "A pipelined FFT processor for word-sequential data," IEEE Transactions on Acoustics, Speech and Signal Processing, vol. 37, pp. 1982 - 1985, 1989
[15] S. He and M. Torkelson, "A new approach to pipeline FFT processor," in Parallel Processing Symposium, Proceedings of IPPS '96, The 10th International Honolulu, HI, USA, 1996, pp. 766 - 770
[16] L. Yang, K. Zhang, H. Liu, J. Huang, and S. Huang, "An Efficient Locally Pipelined FFT Processor," IEEE Transactions on Circuits and Systems II: Express Briefs, vol. 53, pp. 585 - 589, 2006.
[17] S. He and M. Torkelson, "Designing pipeline FFT processor for OFDM (de)modulation," in International Symposium on Signals, Systems, and Electronics (ISSSE'98), 1998, pp. 257 - 262
[18] M. A. B. Nun and M. E. Woodward, "A modular approach to the hardware implementation of digital filters " Radio and Electronic Engineer vol. 46, pp. 393 - 400 1976
[19] A. Peled and B. Liu, Digital signal processing : theory, design, and implementation. New York: Wiley, 1976.
[20] Z. A. M. Sharrif, "Digit slicing architecture for real time digital filters." vol. Ph.D UK: Loughborough University, 1980.
[21] S. A. Samad, A. Ragoub, M. Othman, and Z. A. M. Shariff, "Implementation of a high speed Fast Fourier Transform VLSI chip " Microelectronics Journal, vol. 29, pp. 881-887 1998.
[22] Yazan Samir and T. Rozita, "The Effect Of The Digit Slicing Architecture On The FFT Butterfly," in 10th International Conference on Information Science, Signal Processing and their Applications (ISSPA 2010) Kuala Lumpur, Malaysia, 2010, pp. 802-205.